\documentclass[12pt]{article}
\let\section=\subsection     \let\subsection=\subsubsection                
\usepackage{graphicx}

\begin{document}
\begin{center}
   {\large \bf QCD SPECTROSCOPY AT GSI:}\\[2mm]
   {\large \bf EXOTICA AND CHARMONIA}\\[5mm]
   T.~BARNES \\[5mm]
   {\small \it  
ORNL Physics Division \\
Bldg. 6003, M.S. 6373, Oak Ridge, TN 37831, USA\\
and\\ 
Department of Physics and Astronomy, University of Tennessee\\
Knoxville, TN 37996, USA
\\[8mm] }
\end{center}

\begin{abstract}\noindent
In this talk I give a short summary of the basics of 
conventional and exotic meson
spectroscopy, and consider in particular those issues in the
charmonium and charmonium hybrid sectors which can be addressed by
a future 
antiproton facility at GSI. 
\end{abstract}

\section{QCD and Confinement}

The QCD lagrangian describes the strong interaction in terms of the 
couplings between
the elementary pointlike quark (antiquark) and gluon constituents; 
\begin{equation}
    {\cal L} = 
\sum_q 
\bar \psi_q (i\partial\;\!\!\!\!\! /  - m_q ) \psi_q
-g\; \bar \psi_q (\lambda^a /2) A^a \;\!\!\!\!\!\!\! / \; \psi_q
-{1\over 4}\,  
G_{\mu\nu}^a \,
G_{\mu\nu}^a \ .
\end{equation}
The fundamental interactions between quarks and gluons 
are a QED-like $q\bar qg$ coupling and non-Abelian $g^3$ and $g^4$ 
gluon self-couplings.
At large momentum scales these perturbative interactions 
provide an accurate description
of QCD interactions, and pQCD predictions can be compared 
to the experimentally observed 
cross sections for quark and gluon jet production. At small momentum 
transfers however pQCD becomes inaccurate,
and the dynamics of QCD is instead dominated by the 
nonperturbative phenomenon known 
as {\it confinement}. The gluon self-interactions 
lead to the formation of a ``flux tube"
between color sources, which gives rise to an approximately 
linear potential. Due to this effect only states with zero total
color, ``color-singlets", can exist as physical bound states.
This flux 
tube and the associated asymptotically linear potential are 
clearly evident 
in lattice gauge theory 
simulations with static sources (see Figs.1,2 \cite{Bali1,Bali2}).

\section{Hadrons: Conventional and Exotica}

\subsection{Types of Hadrons}

Hadrons are conventionally classified according to 
which ``valence" basis state in 
Hilbert space is thought to
dominate the 
hadronic state vector. This simple classification 
in terms of pure valence states
provides a useful and surprisingly accurate 
description of most known resonances. 
Of course we should emphasize that this is 
an approximation of unknown accuracy with no clear
justification, and may well be misleading in the description
of unconventional types of hadrons.

\subsection{Quark and Multiquark States}

The physically allowed 
color-singlet states one can form from quarks and antiquarks 
alone are generically
of the form
\begin{equation}
|{\rm color-singlet} \rangle_{n,\bar n} =
|q^n\bar q^{\bar n}\rangle \ ,
\end{equation}
where $mod(n-{\bar n},3)=0$.
The simplest such combinations are
\begin{equation}
\!\!\!\!
|{\rm color-singlet} \rangle_{1,1} 
=
|q\bar q\rangle \ = |{\rm quark\ model\ meson}\rangle \ ,
\end{equation}
\begin{equation}
|{\rm color-singlet} \rangle_{3,0} =
|qqq\rangle \ = |{\rm quark\ model\ baryon}\rangle \ ,
\end{equation}
\begin{equation}
\ \ \ \; |{\rm color-singlet} \rangle_{0,3} =
|{\bar q}{\bar q}{\bar q}\rangle 
=   
|{\rm quark\ model\ antibaryon}\rangle \ .
\end{equation}
The complete spectrum of $q$ and $\bar q$ product basis states 
is formed by 
taking all possible quark types ``flavors" for $q$ and $\bar q$, 
and by allowing the 
states to take on all possible quark spin arrangements and orbital angular
momenta, and finally by allowing excitation of the radial wavefunctions. 
We will discuss the detailed quantum numbers allowed to 
$q\bar q$ mesons in particular in the next section.

These three simplest color-singlet states, 
$|q\bar q\rangle$, 
$|qqq\rangle$ and
$|{\bar q}{\bar q}{\bar q}\rangle$ 
are special in that they are
{\it irreducible}, in other words they
they cannot be partitioned into separate color-singlet
substates. The ``higher" Fock space color singlets in contrast
are {\it reducible}, and need not be realized in nature as isolated
resonances. 
Two examples of such higher Fock space states are
\begin{equation}
|{\rm color-singlet} \rangle_{2,2} =
|q^2{\bar q}^2\rangle \ = |{\rm quark\ model\ baryonium}\rangle \ 
\end{equation}
and 
\begin{equation}
\,\ |{\rm color-singlet} \rangle_{6,0} =
|qqqqqq\rangle \ = |{\rm quark\ model\ dibaryon}\rangle \ .
\end{equation}
The multiquark combination
\begin{equation}
|{\rm color-singlet} \rangle_{4,1} =
|qqqq\bar q\rangle \ 
\end{equation}
is also an allowed color-singlet 
basis state, but has received rather less theoretical attention.

Since these hypothetical baryonia and dibaryons 
have overlap with scattering states of two separate 
$|q\bar q\rangle$ mesons
and two separate $|qqq\rangle$ baryons respectively, 
they can ``decay" without interaction.
It is expected therefore that they
either have extremely broad widths from 
``{\it fall-apart}" into these final states,
or may not be realized in nature as resonances at all. 
This fall-apart problem would be circumvented by a multiquark state with a mass
below all strong decay thresholds, which would therefore be strongly stable. 
Possibilities for
strongly stable multiquark states include the $u^2d^2s^2$ 
``H dibaryon" \cite{Hdibary} and ``heavy-light" 
$Q^2\bar q^2$
clusters, with $Q=c$ or $b$ 
\cite{JMRich}. 

Alternatively, one may 
find quasinuclear bound states of largely undistorted
hadrons that formally lie in multiquark sectors of Hilbert space, 
such as nuclei, hypernuclei, and perhaps $K\bar K$ bound states
\cite{WI}.

\subsection{Quarkonium and $q\bar q$ Quantum Numbers}

Most known mesons are reasonably well described as 
$q\bar q$ (quark-antiquark) bound states. Since quarks have $S=1/2$,
the $q\bar q$ pair can have total spin 
$S_{q\bar q} = 1/2 \otimes 1/2 = 1\oplus 0$. 
The $q\bar q$ orbital
angular momentum $L_{q\bar q}$ can take on any integer value; combining
these 
$L$ and $S$ 
$q\bar q$
angular momenta
gives the allowed total angular momentum $J_{q\bar q}$.
The allowed values are $J=L$ (for $S=0$) and
$J=L+1,L,L-1$ (for $S=1$).
Meson quark model assignments may be specified using 
spectroscopic notation, 
${}^{2{\rm S}+1}{\rm L}_{\rm J}$.
As examples, the $\pi$ is a 
${}^1{\rm S}_0$
state, the 
$J/\psi$ is
${}^3{\rm S}_1$
and the $L=1,S=1,J=2$ $f_2(1270)$ is a
${}^3{\rm P}_2$ $q\bar q$ quark model state. 
Radial excitation may be indicated using a prefactor, thus the
first radially-excited $1^{--}$ $c\bar c$, known as the $\psi(3686)$, is a
$2 {}^3{\rm S}_1$ state.

Spatial parity $P$ and charge-conjugation parity $C$ are exact symmetries of
the QCD lagrangian, and as such are conserved in strong decays. 
In $q\bar q$ states
these quantum numbers are
\begin{equation}
P = (-1)^L
\end{equation}
and 
\begin{equation}
C = (-1)^{L+S} \ .
\end{equation}
A state's 
$J^{PC}$ 
quantum numbers follow directly from these relations;
for example the 
${}^1{\rm S}_0$ $\pi^o$ has
$J^{PC} = 0^{-+}$,
the 
${}^3{\rm S}_1$ $J/\psi$ has
$J^{PC} = 1^{--}$, and the
${}^3{\rm P}_2$ $f_2(1270)$ has
$J^{PC} = 2^{++}$.
As we shall see, sensitive tests of the nature of 
interquark forces are possible given accurate experimental information
on the spectrum of $q\bar q$ states in heavy quark systems;
$c\bar c$ is an especially clear case.

If we complete a table of all possible $J^{PC}$ quantum numbers allowed
to $q\bar q$ states,
we find that certain combinations do not arise.
These 
``$J^{PC}$-exotics" are
$ 0^{--}; 0^{+-}, 1^{-+}, 2^{+-}, 3^{-+} ...$.  
Since relatively low-mass $J^{PC}$-exotics 
are expected in the hybrid meson spectrum, and if discovered would
certainly constitute proof of states beyond the naive $q\bar q$ quark model,
the search for such states is widely regarded as
the most exciting topic in QCD spectroscopy.

\subsection{Glueball and Hybrid States}

Since this subject was covered extensively by other Hirschegg Workshop
speakers in the context of light $(u,d,s,g)$ spectroscopy,
I will be very brief here and proceed to the $c\bar c$ and $c\bar c$-hybrid
states.

One may also form physically allowed (color-singlet) basis states from
pure gluons and from quark, antiquark and gluon product states. 
Hadrons which have such configurations as valence basis states
are known collectively as ``gluonic hadrons".
Color-singlet gluon states (if we neglect quarks) 
form idealized, unmixed ``glueball" resonances. The spectrum of these states
has been studied extensively in LGT (see for example \cite{Morningstar}), and an
impressively detailed theoretical spectrum has been determined.
Unfortunately for experimentalists, LGT predicts only
one glueball below 2~GeV, a scalar at
1.6~GeV, and no $J^{PC}$-exotics are expected below ca. 4~GeV. 
This scalar has been identified with the $f_0(1500)$ seen in 
$p\bar p$ annihilation \cite{f0_1500}, although there are outstanding
problems with clear violation of flavor symmetry in the strong decay
branching fractions of this state; naively one would expect a glueball
to couple symmetrically to all quark flavors. 
This 
flavor-symmetry violation may indicate that
glueball-quarkonium mixing is an important effect \cite{f0_1500}.
This mixing could be important both
in decays and in mass shifts of the observed resonances relative to
LGT predictions; since this is in effect a systematic error for LGT,
it will be very important to quantify.

The relatively narrow width of the 
$f_0(1500)$ glueball candidate is very encouraging for
higher-mass glueball searches; LGT predicts these to be a 
$0^{-+}$
and a
$2^{++}$,
just below 2.5~GeV. Of course identification of these states
will require clarification of the higher quarkonia expected in the same
mass region.   
A future $p\bar p$ machine could make a very useful contribution 
through an
accurate determination of the branching fractions of 
the $f_0(1500)$ and other light glueball and hybrid candidates.

Hybrids are the most experimentally attractive of the 
anticipated non-$q\bar q$ resonances, because their valence
$|q\bar q g\rangle $ basis states span complete flavor nonets
(hence hybrids have a much richer spectrum of states than glueballs)
and the lowest-lying hybrid multiplet is expected to contain 
exotic quantum numbers. 
{\it All} $J^{PC}$ combinations can be formed from 
$|q\bar q g\rangle $ states, so any $J^{PC}$-exotic might {\it a priori} 
be a hybrid meson candidate. LGT has recently contributed several estimates
of light exotic hybrid meson masses \cite{CM,ccH_LGT}, 
and at present it appears that the
combination $J^{PC}=1^{-+}$ is the lightest exotic, with a mass of
$M(1^{-+})\approx 2.0$~GeV in the $(u,d)$ $q\bar q$ flavor sector. This is
consistent with estimates using the flux-tube model \cite{hyft}, 
and rather heavier than bag model results, which favored $M(1^{-+})\sim 1.5$~GeV.

We can expect to identify the $J^{PC}$-exotic hybrids rather easily,
provided that we study their favored decay modes. The expectation of both
flux tube \cite{hyft} 
and 
constituent gluon \cite{hycg} 
models is that
light hybrids should decay preferentially to pairs of $q\bar q$ mesons in
which one has a unit of orbital excitation, the so-called ``S+P" modes.
For the lightest $I=1$, $J^{PC}=1^{-+}$ hybrid these modes are
$\pi f_1$ and $\pi b_1$, which are rather difficult to reconstruct
and so had not been investigated carefully before the recent interest in
hybrid mesons.

As a caution we note that the only 
two light exotic hybrid candidates, the
$\pi_1(1400)$ and $\pi_1(1600)$, both have these
$I=1$, $J^{PC}=1^{-+}$ quantum numbers, but lie ca. 500~MeV below 
the LGT and flux-tube mass 
predictions
and apparently decay strongly to the 
S+S modes $\eta\pi$ and $\rho\pi$, which are forbidden to hybrids in the
flux-tube decay model.
Evidently
these theoretical expectations for hybrids should not be regarded as more 
than tentative
guidelines at present. We can of course expect a systematic improvement in 
LGT predictions as algorithms and computer performance improve.

\section{Charmonium and $c\bar c$ Hybrids}

\subsection{Charmonium, Theory and Experiment}

There are 11 known
$c\bar c$ resonances; the spectrum is shown in Fig.3.
(A possible 12{\it th} $c\bar c$ state, 
a candidate for the anticipated narrow $^3$D$_2$,  
was reported 
in $J/\psi \pi^+\pi^-$ 
at 3.836~GeV 
by 
the E705 Collaboration \cite{E705}. This effect has
a rather low $2.8\sigma$ statistical significance, and needs 
confirmation.) 
There is a predominance of 
$J^{PC}=1^{--}$ 
vector states simply 
because most of these resonances were found at $e^+e^-$
machines, which form only $J^{PC}=1^{--}$ states in s-channel. The remaining 
states in the 1P multiplet and the $^1$S$_0$ spin-singlet $\eta_c(2980)$
were found in radiative transitions from the 1S and 2S vectors, with the
single exception of the $h_c(3526)$; this $J^{PC}=1^{-+}$ state has been
observed only in $p\bar p$ annihilation.

Since charmonium is only quasirelativistic, one can expect that the
level of configuration mixing is much reduced 
relative to light mesons
(the gluon emission 
amplitude is $\propto (v_q/c_q)$), so that one may be able to clearly
identify the separate effects of confinement and gluon exchange 
($|c\bar c\rangle \leftrightarrow |c\bar c g\rangle $ mixing) in the 
$c\bar c$ spectrum. 
Since spin-dependent forces 
do not 
appear in the effective interquark interaction until
$O(v^2/c^2)$ in the quark momenta, 
we might expect that a naive
zeroth-order static potential that incorporated the
OGE color-Coulomb potential and linear
confinement might give a reasonable approximation to the observed
$c\bar c$ spectrum \cite{GI}.
We can test this by assuming the potential
\begin{equation}
V_{c\bar c}(r) = -{4\over 3} {\alpha_s\over r} + br 
\end{equation}
and solving the nonrelativistic Schr\"odinger equation for bound states 
in this potential.
Inspection of the experimental $c\bar c$ spectrum (Fig.3) 
suggests 1S, 1P and
2S multiplets with spin-averaged masses of ca. 
3.07~GeV,
3.52~GeV and 
3.67~GeV, respectively. If we use these as input to fix the three potential
model parameters $\alpha_s, b$ and the charm quark mass 
$m_c$, we find 
$\alpha_s = 0.510$,
$ b = 0.152$~GeV$^2$ and
$m_c = 1.450$~GeV.
Most of the remaining $c\bar c$ levels predicted to lie below 4.6~GeV
(1S..4S,1P..3P,1D..3D,1F,2F,1G,2G and 1H) are shown in Fig.4,
together with the experimental spectrum.
With a multiplicity of 2 for S-states 
($^1$S$_0$,$^3$S$_1$) and 4 for higher-$L$ states, this model 
predicts 52 independent $c\bar c$ states below 4.6~GeV.
The 
proximity of the experimental masses to the predicted
radial and orbital levels 
in Fig.4 confirms
that the simple description of $c\bar c$ states as nonrelativistic 
fermions in a
Coulomb-plus-linear potential is a reasonable first approximation.

The level splittings within an orbitally-excited
multiplet such as 
1P provide more sensitive tests of the nature of
interquark forces. Assuming that the short-ranged force is 
due to one-gluon exchange (OGE), we expect the spin-dependent
forces to be reasonably well described by the 
Breit-Fermi Hamiltonian, which follows from an
$O(v^2/c^2)$ 
expansion of the OGE T-matrix.
This Breit-Fermi interaction, which is familiar from atomic physics,
has spin-spin, spin-orbit and tensor terms, and 
for equal-mass quarks and antiquarks ($m_q=m_{\bar q}=m$) is explicitly
\begin{displaymath}
H^{OGE}_{Breit-Fermi} = \hskip 8cm
\end{displaymath}
\begin{equation}
{32\pi 
\alpha_s   
\over  9 m^2 }\;
\vec S_q \cdot \vec S_{\bar q} \;
\;
\delta(\vec x\, )
+
{2 
\alpha_s   
\over  m^2 r^3 }\;
\vec L_{q\bar q} \cdot \vec S_{q\bar q} \;
\;
+
{4\pi 
\alpha_s   
\over  m^2 r^3 }\;
( \vec S_q\cdot \hat r \,   \vec S_{\bar q}\cdot \hat r 
- {1\over 3}\, \vec S_q \cdot \vec S_{\bar q} )
\;
\ .
\end{equation}
\noindent
There are also spin-dependent terms which change the relative positions of 
multiplets by $O(v^2/c^2)$ but do not separate levels within a multiplet.

Several very characteristic features of this OGE interaction are immediately
evident. First, since the spin-spin interaction is a contact term, it has
no effect on orbitally-excited states. Thus the spin-singlet state
$(S=0,J=L)$ 
is predicted to be degenerate with the multiplicity-weighted
``center-of-gravity" of the
spin-triplet states
$(S=1,J=L+1,L,L-1)$. (The spin-orbit and tensor mass shifts 
give zero when weighted
by multiplicity.) Thus in the 1P multiplet we predict that
the 
${}^1$P$_1$ $h_c$ spin singlet should have a mass of
\begin{equation}
M({}^1{\rm P}_1)\bigg|_{thy. (OGE)} =
{5\over 9}\, M({}^3{\rm P}_2) +
{3\over 9}\, M({}^3{\rm P}_1) +
{1\over 9}\, M({}^3{\rm P}_0) 
= 3525.27(0.12) \ {\rm MeV}.
\end{equation}
This relation is very well satisfied by the
experimental candidate 
$h_c(3526)$ 
reported 
by the Fermilab collaboration E760/835
in $p\bar p$
annihilation \cite{PDG_hc}; it has a mass of 
\begin{equation}
M({}^1{\rm P}_1)\bigg|_{expt.} 
= 3526.14(0.24)  \ {\rm MeV}.
\end{equation}
The agreement is not expected to be exact because this is an
$O(v^2/c^2)$ derivation, and makes additional approximations
such as assuming pure $c\bar c$ states and {\it only} OGE 
at small $r$.

This result is often cited as a sensitive test of the 
Lorentz nature of the confining interaction. {\it A priori}
one might have assumed that the confining interaction
couples to the color charge density 
$\psi ^\dagger \psi 
=\bar \psi \gamma_0 \psi$, so that the complete quark-antiquark
interaction is of the same form as the QED Coulomb interaction,
$\gamma_0 \otimes \gamma_0$. 
This was assumed in the original
Cornell model, and is still advocated 
by Swanson and Szczepaniak \cite{SS}.  
Alternatively the confining interaction
might couple to the Lorentz scalar density $\bar \psi \psi$, so that the
complete interaction transforms as
$I \otimes I$. These two possibilities may be distinguished by
the $O(v^2/c^2)$ spin-dependent Hamiltonian. The general result for the
spin-spin $c\bar c$ interaction due to a 
$\gamma_0 \otimes \gamma_0$ potential $V(r)$ is
\begin{equation}
H^{spin-spin}_{vector} = 
+
{2
\over   3 m_c^2  }\;
{\nabla}^2\, V(r) \;
\vec S_q \cdot \vec S_{\bar q} 
\ .
\end{equation}
With a {\it vector} linear confining interaction $V(r) = b_v r$ this becomes
\begin{equation}
H^{spin-spin}_{vector\ conft.} = 
+
{4 b_v
\over   3 m_c^2 r }\;
\vec S_q \cdot \vec S_{\bar q} 
\ .
\end{equation}
This vector confinement would displace the 
$^1$P$_1$ 
$h_c$ state upwards in mass from the 
$^3$P$_{\rm J}$  c.o.g. by 
$(4b_v/3m_c^2)\; \langle {\rm 1P}|\, r^{-1}|{\rm 1P}\rangle
\approx 30$~MeV; since these energies
are actually equal to within about 1~MeV, this is a very strong
argument in favor of scalar over vector confinement. 
Since the
$h_c(3526)$ state is not very well established 
experimentally, and has been seen only in $p\bar p$ annihilation, 
confirmation of this 
$^1$P$_1$
state and a precise mass determination  
will be a very important exercise for GSI.

In addition to this OGE interaction there is an inverted spin-orbit 
interaction due to the linear scalar confining potential, which is
given by
\begin{equation}
H^{spin-orbit}_{scalar\ conft.} = 
-
{\pi b
\over   2 m_c^2 r }\;
\vec L_{q\bar q} \cdot \vec S_{q\bar q} \;
\ .
\end{equation}

The effect of incorporating these additional spin-dependent terms 
as first order perturbations is shown in Fig.5. Note that this is
{\it not} a fit to the observed multiplet
splittings, rather these splittings follow from the
$\alpha_s, b$ and $m_c$ that fit the 1S, 1P and 2S multiplet
centers of gravity (Fig.4). 
As a qualitative description of the relative positions
and scale of splittings within the multiplet this model is 
evidently quite successful. The relative splitting 
$
( {}^3{\rm P}_2 - {}^3{\rm P}_1 ) / 
( {}^3{\rm P}_2 - {}^3{\rm P}_0 )
$
is an especially interesting quantity, since the 
wavefunction uncertainties approximately cancel and one can see the
effects of the negative (scalar confinement) spin-orbit and (small)
OGE tensor terms. The observed ratio is rather close to 
theoretical expectations from OGE and linear scalar confinement. 
Similar tests in the 2P and especially 1D multiplets would be quite
interesting at GSI, since the negative scalar spin-orbit term is
longer ranged than the OGE spin-orbit, so we expect to see considerable
narrowing of the multiplet splittings with increasing orbital and radial
excitations. Complete inversion is predicted with increasing $L$, 
but other effects such as configuration mixing and coupling to open charm
states may mask this interesting effect. Accurate mass determinations
of many conventional $c\bar c$ states above open charm threshold
will be very useful for theorists trying to quantify the various mass shifts.

Identification of the $c\bar c$ spectrum above the open-charm threshold
at 3.73~GeV will be interesting for tests of decay models, spectroscopy models,
and also because these states are a ``background" which might 
otherwise be confused with
charmonium hybrids, charm meson molecules, or other unusual states.
First, the two 1D states 
${}^1{\rm D}_2$ ($2^{-+}$) and ${}^3{\rm D}_2$ ($2^{--}$) 
are especially interesting because they cannot decay to $DD$, and hence
should be relatively narrow. 
($DD$ is of course an abbreviation for $D\bar D$ in this context.)
Studies of the relative 
branching fractions of other
higher-mass $c\bar c$ states 
to the presumably dominant open-charm modes 
($DD,D^*D,D^*D^*,D_sD,D^*_JD,...$) will be an extremely interesting
contribution to our understanding of strong QCD physics. 
Theorists usually treat
open-flavor strong decays 
using the ``$^3$P$_0$ decay model" or
one of its variants such as the flux-tube model. This type of decay model,
which predates QCD, 
describes open-flavor decays as due to $q\bar q$ pair creation 
with vacuum ($^3$P$_0$) quantum numbers. Just why this model works is unclear,
and the evidence supporting it is rather meagre; 
the classic tests are the 
$D/S$ 
amplitude ratios
in the two decays
$b_1\to \omega\pi$ and
$a_1\to \rho\pi$ \cite{HQ}.
Since so much of hadron spectroscopy makes use of this decay model 
(the weak $\pi N$ solution of the ``missing baryon" problem and the S+P hybrid 
signature are two examples), it is very important to test it
using a wide range of resonance quantum numbers and final states.
The higher-mass $c\bar c$ states will be very useful in this regard. Calculations
of the 
open-charm
branching fractions of higher $c\bar c$ states have previously been reported
using
the $^3$P$_0$ model \cite{ccbar}, and these show interesting dependence on
the nodal structure of the radial $c\bar c$ wavefunctions.
An experimental determination of the strong decay amplitudes of the accessible
higher-mass $c\bar c$ 
states would allow an extremely interesting test of this widely used
but inadequately tested strong decay model.

The little that is known about strong decays of the higher-mass 
$c\bar c$ states already
includes a famous puzzle; 
the $\psi(4040)$, which has a mass consistent with a 3S state
(Fig.4), purportedly has relative branching fractions 
of $D^*D^* >> D^*D >> DD$, despite 
the fact that the $D^*D^*$ mode has essentially 
no phase space! This led 
to speculations that the $\psi(4040)$ might be a $D^*D^*$ molecule
\cite{molec}, or that it may be the expected 3 $^3$S$_1$ 
$c\bar c$ state, but with nearby decay
amplitude zeros that lead to these anomalous 
branching fractions \cite{ccbar}. Since we hope to use branching fractions
to characterize states,
an accurate test of these
and other strong branching fractions would 
clearly be a first priority at a new $p\bar p$
charmonium facility.

\subsection{Charmonium Hybrids}

The charmonium system is an excellent laboratory for the study of
nonperturbative QCD effects such as confinement and gluonic excitations.
It has the advantage of being quasirelativistic; the adiabatic
$c\bar c$ potential is clearly evident in the spectrum of states,
but the $O(v/c)$ $|c\bar c g\rangle$ gluonic
configuation mixing is sufficiently large to be accurately determined
and compared with model predictions, for example in
the $O(v^2/c^2)$ spin-dependent multiplet splittings of the 1P states.  
The simplicity of the known
$c\bar c$ spectrum suggests that it may be straightforward to identify
relatively unmixed charmonium hybrids as ``extra" charmonium states
through a more complete determination of 
the experimental 
spectrum. Of course the identification of complete hybrid multiplets,
especially $J^{PC}$ exotics, would be a crucial contribution to
our understanding of the dynamics of gluonic excitations.
In the charmonium system these states may be narrow enough to make this
a feasible experimental program.

Recent theoretical advances have simplified the problem of searching for
hybrid charmonium considerably. Previous model estimates of the mass of
the lowest 
hybrid charmonium multiplet 
varied over a rather wide range, ca. 4.0-4.5~GeV. 
(For a review of this earlier work see Ref.\cite{cc_tcf}.)
With the development of lattice NRQCD we now have 
lattice results for the masses of exotic $c\bar c$- and $b\bar b$-hybrids
that report very small statistical errors of ca. 10~MeV. (The systematic 
uncertainties are not yet known but might be ca. 50~MeV, and will be
estimated in subsequent work.) As one example, the CP-PACS collaboration 
\cite{ccH_LGT} quote masses for the $1^{-+}$ 
(expected to be the
lightest exotic) 
$c\bar c$ and $b\bar b$ hybrid states 
of
\begin{equation}
M_{c\bar c H}(1^{-+}) = 
M_{\ c\bar c}(1S)  + 1.323(13)\ {\rm GeV} \approx 4.39\ {\rm GeV} \ 
\end{equation}
and
\begin{equation}
M_{b\bar b H}(1^{-+}) = 
M_{\ b\bar b}(1S)  + 1.542(8)\ {\rm GeV} \approx 10.99\ {\rm GeV} \ .
\end{equation}
(I assume multiplicity-weighted 1S masses of 3.07~GeV 
for $c\bar c$ and 9.45~GeV 
for $b\bar b$.)
Thus we have a presumably accurate lattice estimate of the mass of 
the lightest $c\bar c$ hybrid multiplet,
$\approx 4.4$~GeV. 
The precise mass of the lightest hybrid multiplet
has previously been of great interest because of the flux-tube model
prediction that S+P modes should be strongly favored for hybrids;
if $c\bar c$-hybrids were below the S+P threshold of 
$M(D) + M(D^*_J)\approx 4.25$~GeV, one might have anticipated relatively
narrow states. With the NRQCD lattice results it now appears
that S+P modes are indeed open, so $c\bar c$-hybrids need
not be anomalously narrow. In any case the observation of important
$\pi\eta$ and $\pi\rho$ modes for the hybrid candidates
$\pi_1(1400)$ 
and
$\pi_1(1600)$ suggests that experiment may not support this selection
rule as an especially rigorous one; the simple S+S modes $DD$, $DD^*$ and
$D^*D^*$ may well be the dominant hybrid modes. It will be very important
experimentally to search all allowed quasi-two-body open charm modes
for these states. 
Just as with the conventional $c\bar c$ states, much is speculated but very 
little is known about
open-flavor strong branching fractions of hybrids.

\section{Summary and Conclusions}

I would like to thank H.Koch, W.Norenberg and the organisers of the
Hirschegg Workshop for their support and their
kind invitation to present this material, and
for the opportunity to discuss hadron spectroscopy with my fellow 
participants.
This research was sponsored in part by the U. S. Department of Energy
under contract DE-AC05-00OR22725, 
managed by UT-Battelle, LLC.

\vskip 1cm

\begin{center}
\includegraphics[width=6cm,height=6cm,angle=00]{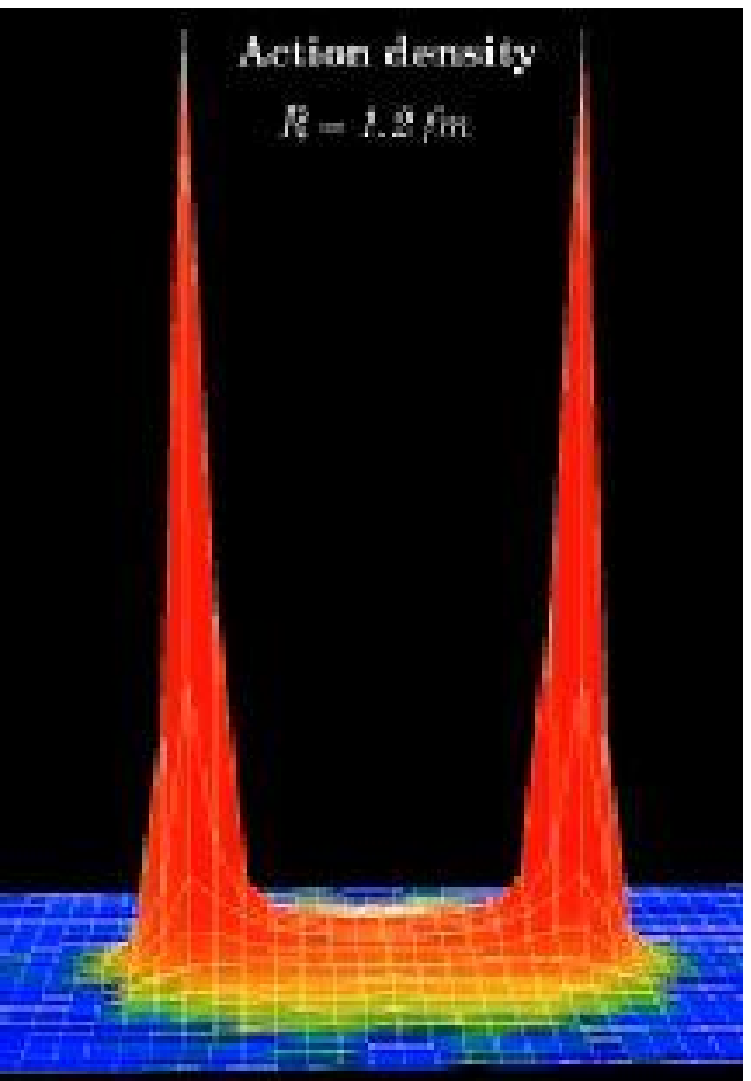}\\
\parbox{14cm}
\centerline{\footnotesize \\  
Fig.~1: The gluonic flux tube between static quark sources 
(Bali et al. \cite{Bali1}). }
\end{center}

\begin{center}
\includegraphics[width=8cm,height=8cm,angle=00]{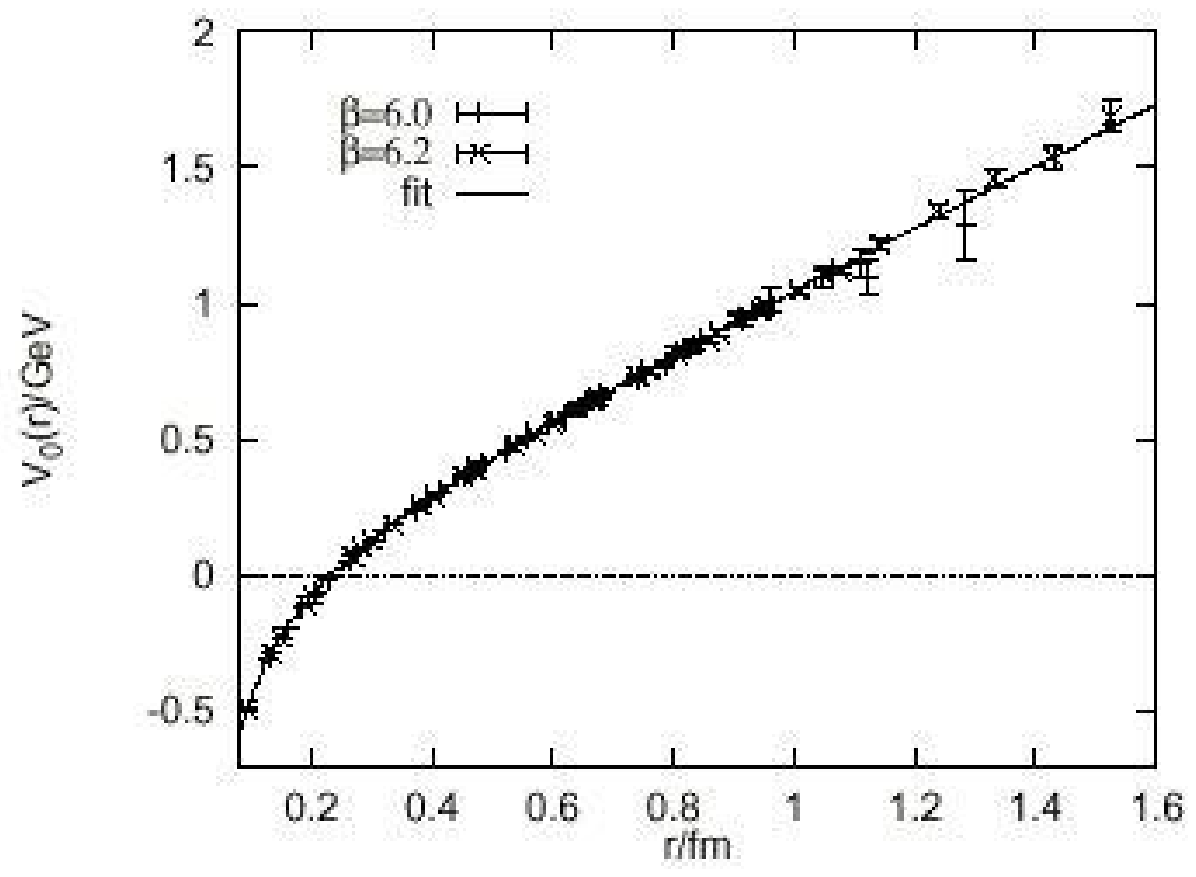}\\
\parbox{16cm}
\centerline{\footnotesize \\
Fig.~2: An LGT determination of the interquark
potential (Bali et al. \cite{Bali2}).}
\end{center}

\begin{center}
\includegraphics[width=8cm,height=10cm,angle=-90]{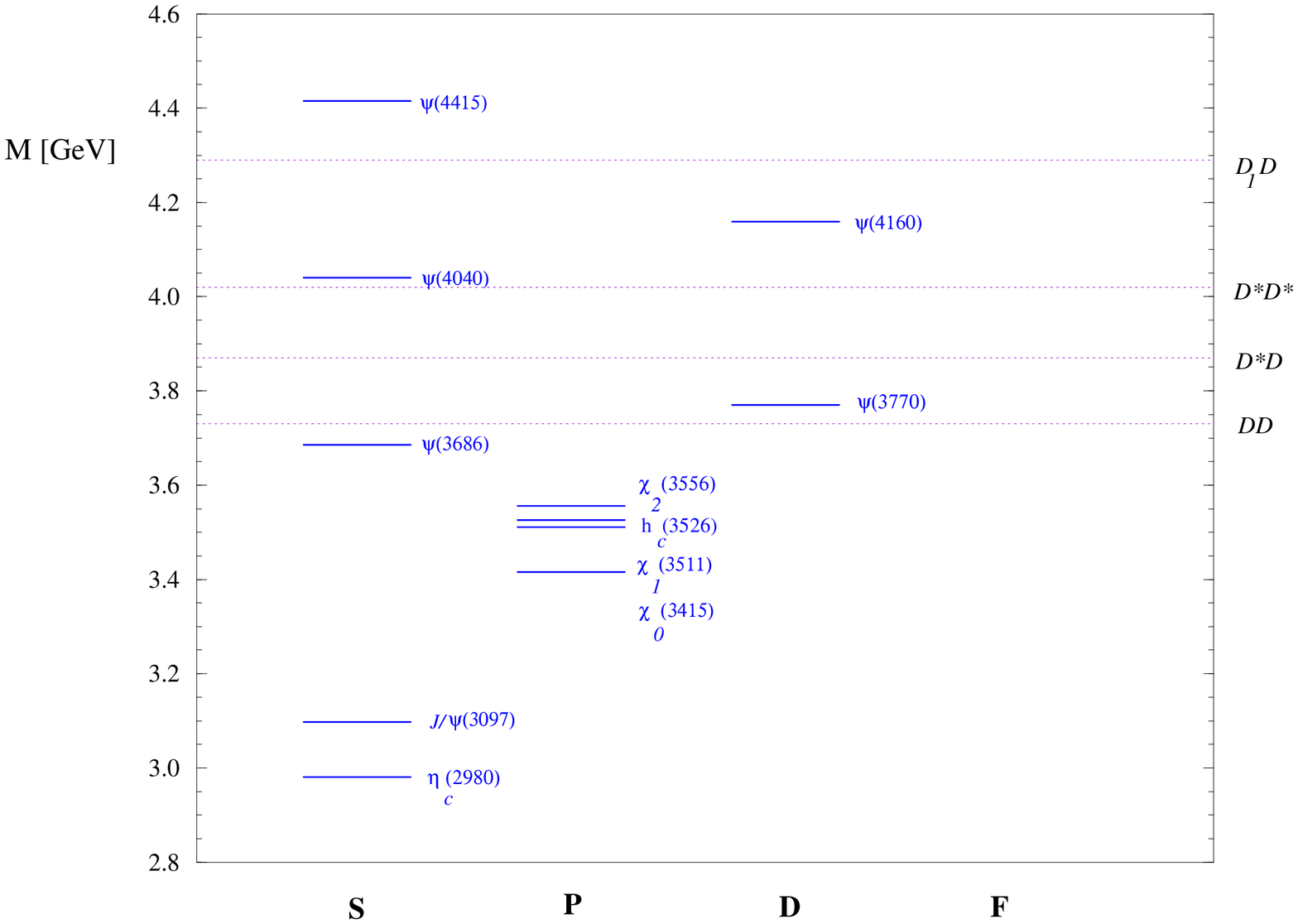}\\
\parbox{14cm}
\centerline{\footnotesize \\
Fig.~3: The 11 known $c\bar c$ states. 
All 
have $J^{PC}=1^{--}$
except the 1P
multiplet near 3.5~GeV and the $J^{PC}=0^{-+}$ $\eta_c(2980)$. 
Some open
charm thresholds are also shown.
}
\end{center}

\begin{center}
\includegraphics[width=8cm,height=10cm,angle=-90]{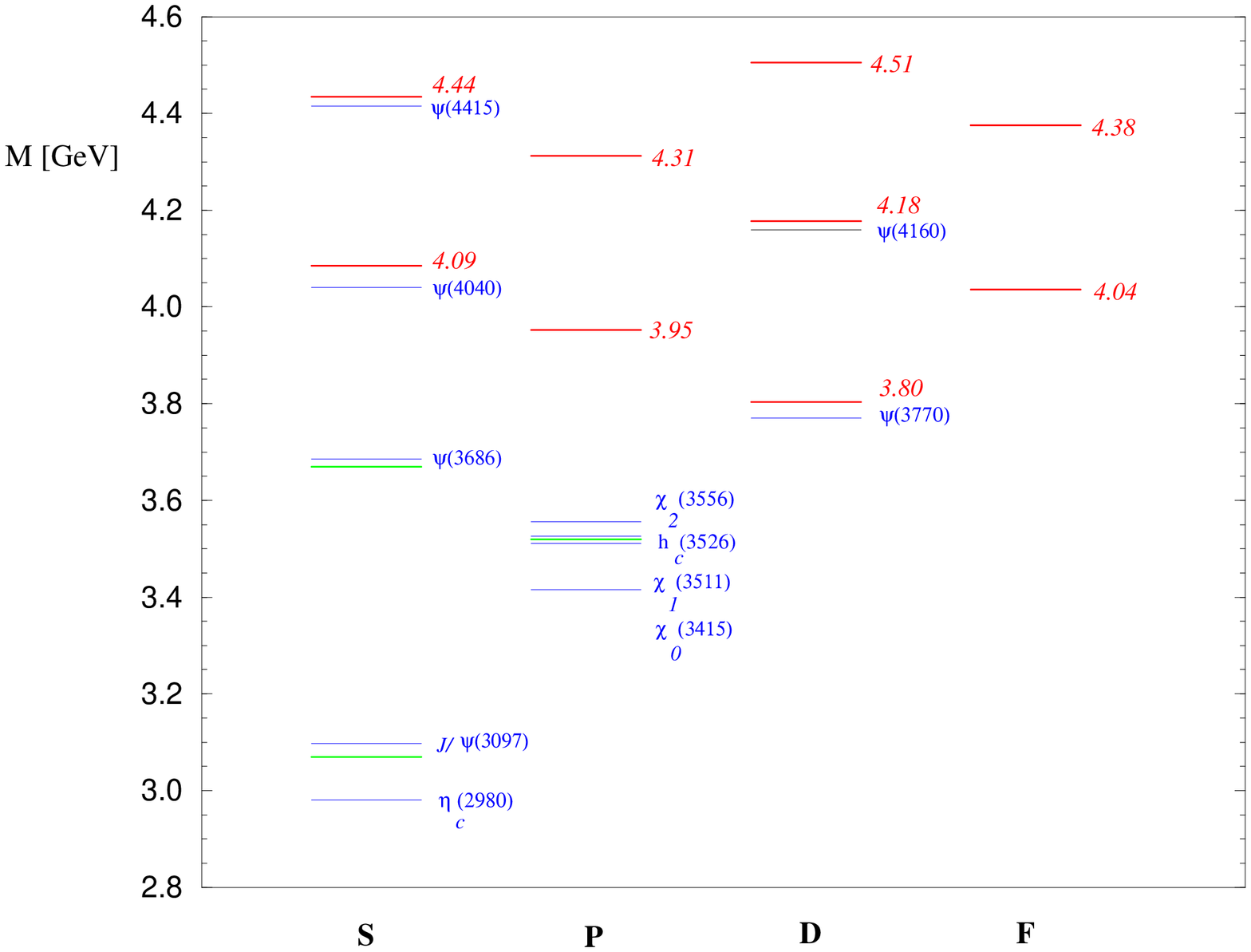}\\
\parbox{14cm}
\centerline{\footnotesize \\
Fig.~4: A comparison of theory (red, 52 states) 
and experiment (blue, 11 states) for $c\bar c$ levels below 
4.6~GeV. 
Mean 1S, 1P and 2S ({\it estm.}) levels (green)  
were input.
Predicted levels not shown are 1G(4.24), 2G(4.56) and 1H(4.43).}
\end{center}

\begin{center}
\includegraphics[width=8cm,height=10cm,angle=-90]{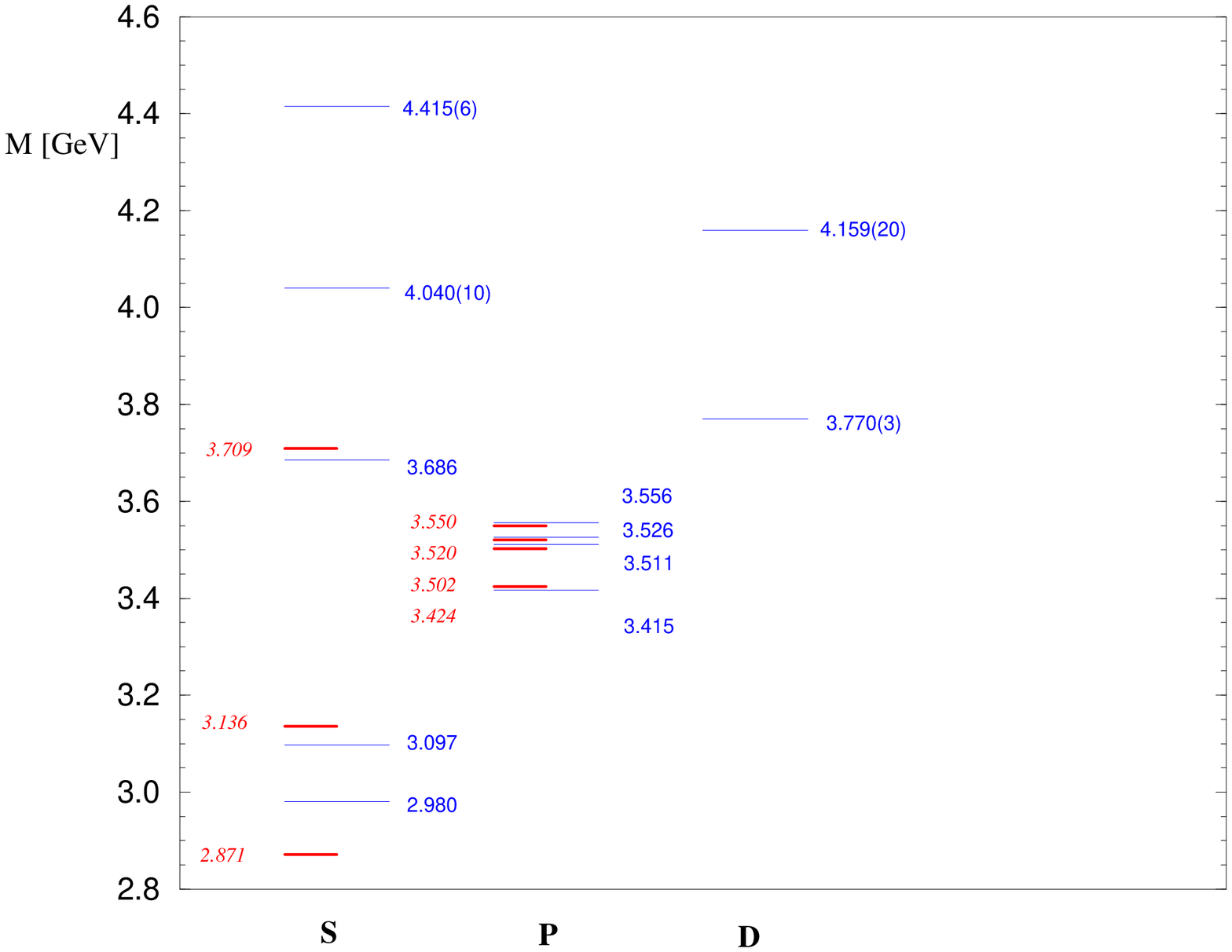}\\
\parbox{14cm}
\centerline{\footnotesize \\
Fig.~5: Spin-dependent splittings from OGE and linear scalar confinement.
Theory (red) is compared to experiment (blue).
This is 
{\it not} a new fit; the parameters are those of Fig.4.}
\end{center}

\end{document}